# Lateral Optical Force on Chiral Particles Near a Surface

S. B. Wang and C. T. Chan*

*Department of Physics and Institute for Advanced Study,
The Hong Kong University of Science and Technology,
Clear Water Bay, Hong Kong, P. R. China*

* Email: phchan@ust.hk

## Abstract

Light can exert radiation pressure on any object it encounters and that resulting optical force can be used to manipulate particles. It is commonly assumed that light should move a particle forward and indeed an incident plane wave with a photon momentum $\hbar\mathbf{k}$ can only push any particle, independent of its properties, in the direction of $\mathbf{k}$. Here we demonstrate using full-wave simulations that an anomalous lateral force can be induced in a direction perpendicular to that of the incident photon momentum if a chiral particle is placed above a substrate that does not break any left-right symmetry. Analytical theory shows that the lateral force emerges from the coupling between structural chirality (the handedness of the chiral particle) and the light reflected from the substrate surface. Such coupling induces a sideway force that pushes chiral particles with opposite handedness in opposite directions.

Electromagnetic (EM) waves carry linear momentum as each photon has a linear momentum of $\hbar k$ in the direction of propagation. Circularly polarized light carries angular momentum due to the intrinsic spin angular momentum (SAM) of photons[1-6]. When light is scattered or absorbed by a particle, the transfer of momentum can cause the particle to move. Thus light can be used to manipulate particles[7-25]. Light will push a particle in the direction of light propagation (as illustrated in Fig. 1a) irrespective of the polarization of light and irrespective of the particle's own properties, even if it has chirality (Fig. 1c, e), as long as we have a plane wave incidence (i.e. a well-defined **k**). Let us now consider the configuration shown in Fig. 1b, which shows a particle placed close to a surface made of an ordinary material (e.g. Au or Si or silica). In this case one might still expect the particle to be pushed in the direction of light propagation, as the surface does not break the left-right symmetry. We will show that this is true only if the particle is non-chiral. If the particle has chiral character, however, it will experience an additional lateral force in a direction that depends on its own chirality as shown in Fig. 1d, f. This counter-intuitive force comes from a lateral radiation pressure and an optical spin density force that couple the chirality of the particle to both the lateral linear momentum and SAM generated by the scattered wave of the chiral particle. The time-averaged SAM densities, defined as $\langle \mathbf{L}_e \rangle = \varepsilon_0/(4\omega i)(\mathbf{E}\times\mathbf{E}^*)$ and $\langle \mathbf{L}_m \rangle = \mu_0/(4\omega i)(\mathbf{H}\times\mathbf{H}^*)$ respectively for the electric and magnetic part, are associated with the polarization of light [26, 27]. The lateral force could move particles with chirality of different signs in different directions as shown in Fig. 1d, f. A good example of a chiral particle is the helix shown in Fig. 1g. Interactions between such kind of chiral objects and EM waves have been extensively studied [28] and are shown to give rise to interesting phenomena such as polarization conversion [29-34], photonic topological edge states [35] and negative-refractive metamaterials [36-39]. We will show that the coupling of EM near field and structural chirality (the handedness of a chiral particle) will induce an anomalous lateral force that pushes the particle sideways rather than forward.

We first show the full-wave numerical results of the optical forces acting on helical gold particles induced by a linearly polarized plane wave. We then consider a simpler system consisting of a model chiral sphere placed above a substrate. To reveal the underlying physics and trace the origin of this intriguing phenomenon, we further simplify the configuration by considering a dipolar chiral particle above a substrate, in which case the problem can be analytically addressed.

## Results

**Lateral optical force on a gold helix placed above a substrate.** Consider a gold helix with inner radius $r = 50$ nm, outer radius $R = 150$ nm and pitch $p = 300$ nm, as shown in Fig. 1g. We are interested in the optical force acting on such a particle induced by a linearly polarized plane wave of the form $\mathbf{E}_{inc} = \hat{z} E_0 \exp(ik_0 x - i\omega t)$. The particle is located above a substrate with a gap-distance of 10 nm and its axis is along the $x$ direction (see Figure 1d, f). The substrate has the dimensions of $l \times w \times t$ (see Fig. 1b) and it can be metallic (e.g. gold) or dielectric.

For an isolated helical particle, the scattering force induced by the plane wave will push it in the direction of $\mathbf{k}_0$ independent of the handedness of the helix (see Fig. 1c, e). However, if the helix is placed above a substrate (as shown in Fig. 1d, f), an additional lateral force ($F_y$) will emerge and push it sideways. The lateral force acting on a left-handed (LH) and a right-handed (RH) particle takes opposite signs. The optical force is calculated numerically by the Maxwell stress tensor method (see Methods). Figure 2a shows the lateral forces acting on an LH (blue lines) gold helix and an RH (red lines) one consisting of four pitches when they are placed above a gold (described by a Drude model, see Numerical simulation) substrate (solid circles) and a dielectric substrate with $\varepsilon_d = 2.5$ (hollow triangles). The lateral forces are evident in a wide range of frequencies, where local resonances associated with the geometric shapes of the particles result in some oscillations. It is important to note that the lateral force takes opposite signs for the LH and RH helices. The existence of a lateral force in the case of a perfect-electric-conductor substrate is also examined and similar results are obtained (see Supplementary Fig. 1).

Figure 2b shows the dependence of the lateral force on the number of pitches there are in the helix at $f = 490$ THz. The magnitudes of the lateral forces in the cases of a gold substrate (solid circles) increase with the pitch number, which indicates that the magnitude of the lateral force increases with the chirality. In the absence of a substrate, there is a small residual lateral force (solid triangles) due to the end effect and this residual lateral force decreases as the pitch number increases (the residual lateral force is zero if the particle is symmetric on the *yz*-plane). Figure 2c shows that the magnitude of the lateral force decreases when the gap distance between the particle and the gold substrate is increased, indicating that the force is due to the coupling between the particle and the substrate. Figure 2d shows the lateral force as a function of the thickness *t* of the dielectric substrate, where the frequency is set at $f = 380$ THz. The lateral force undergoes oscillations and we will see later that the oscillations are due to the Fabry-Perot resonances associated with the dielectric substrate. Note that we do not need to consider the effect of thickness in the case of the gold substrate due to the finite penetration depth of fields.

The numerical results are rather counter-intuitive. The standalone helix (i.e., without a substrate), whether it be LH or RH and assuming it is long enough so that the end effect can be ignored, scatters the same amount of light to the left (+*y*) and to the right (-*y*) and hence light cannot push it sideways. However, when the helix is placed above a substrate, it experiences a lateral force in the *y* direction. To see pictorially how this happens, we examine the magnetic field ($H_x$) patterns for both an RH (Fig. 3a) and an LH (Fig. 3b) helix in the case of the gold substrate at $f = 420$ THz. Figure 3a, b shows that the scattered field is asymmetrically distributed. The surface waves propagate predominately in the +*y* direction in the case of the RH particle while they propagate in the –*y* direction in the case of the LH particle. These field patterns show that the helix-above-substrate configuration produces asymmetric scattering and hence a lateral force, although individually the helix and the substrate produce symmetric scattering.

**Lateral optical force on a chiral sphere placed above a substrate.** The numerical results on the gold helix demonstrates the existence of the lateral force. We now show that the effect can be observed even for a chiral sphere described by simple constitutive relations. Consider again the configuration shown in Fig. 1b, where the spherical particle with radius $r = 75$ nm is now made of a model bi-isotropic chiral material described by the constitutive relations [40-44]:

$$\begin{bmatrix} \mathbf{D} \\ \mathbf{B} \end{bmatrix} = \begin{bmatrix} \varepsilon_r \varepsilon_0 & i\kappa/c \\ -i\kappa/c & \mu_r \mu_0 \end{bmatrix} \begin{bmatrix} \mathbf{E} \\ \mathbf{H} \end{bmatrix}, \quad (1)$$

where $\varepsilon_r, \mu_r$ are the relative permittivity and permeability of the material, respectively; $\kappa$ is the chirality parameter and takes a real value here; $\varepsilon_0, \mu_0$, $c$ are the permittivity, permeability and the speed of light in vacuum. The dispersion for a plane wave in such a medium is $k_\pm = (\sqrt{\varepsilon_r \mu_r} \pm \kappa)\omega/c$, corresponding to circularly polarized states of light. The gap distance between the sphere ($\varepsilon_r = 2, \mu_r = 1$) and the surface of the gold substrate is 10 nm. The particle is under the excitation of the same linearly polarized plane wave defined before. The wavelength is set at $\lambda = 600$ nm. Figure 4a, b shows the time-averaged Poynting vector for the cases with and without the gold substrate, computed by the numerical solver COMSOL Multiphysics[45]. For the isolated particle (Fig. 4a), the Poynting vector has a rotation pattern due to the chiral property of the sphere and the scattered energy is symmetric with respect to –*y* and +*y*. Such rotating Poynting vector pattern can also be found in the case of the gold helix (see Supplementary Fig. 2). However, in the presence of the gold substrate, the "handedness" of the Poynting vector in the near-field region results in an asymmetric pattern as shown in Fig. 4b and hence a lateral force. Figure 4d, e shows the flow of the time-averaged electric SAM[$\langle \mathbf{L}_e \rangle = \varepsilon_0 / (4\omega i)(\mathbf{E} \times \mathbf{E}^*)$]. Similar to the Poynting vector distribution, the computed SAM has a symmetric *y*-component in the isolated sphere case (Fig. 4d). However, in the presence of a gold substrate shown in Fig. 4e, the SAM tends to point in -*y* direction. We will show (through analytical theory) that the asymmetric distributions of Poynting vector and SAM are the consequences of particle-substrate coupling and are closely related to the origin of the lateral force. The blue line in Fig. 4g shows the numerically evaluated lateral force as a function of the chirality parameter $\kappa$ for the case with the gold substrate. The magnitude of the force increases with the magnitude of the chirality. If the chirality is positive, the force is negative, and vice versa. The force vanishes when the medium is non-chiral ($\kappa = 0$). The chirality determines the direction of EM energy coupling and this property can be used to realize uni-directional excitation of surface waves [46-48].

The results shown in Fig. 4a-f also provide an intuitive understanding of the lateral force. From a symmetry point of view, the time-averaged Poynting vectors corresponding to the total fields for a standalone chiral/spiral particle under linearly polarized light excitation must satisfy $\langle \mathbf{S}(y,z) \rangle = -\langle \mathbf{S}(-y,-z) \rangle$ due to the rotational symmetry of the system. This can be seen by examining the numerical results in Fig. 4a. This implies that the total momentum flux scattered by the chiral object to the right (-*y*) and left (+*y*) should be the same for a standalone particle, which in turn implies that there cannot be a net force in the lateral (*y*) direction. However, if we break the $z \to -z$ symmetry of the environment by putting a substrate underneath the particle, there is no symmetry requirement for $|\langle \mathbf{S}(y,z)\rangle|$ and $|\langle \mathbf{S}(-y,-z)\rangle|$ to be equal. This is indeed the case for the numerically calculated Poynting vector pattern as shown in Figure 4b. A lateral force ($F_y$) can now exist as the total photon momentum scattered to the left and right are not balanced, and the magnitude of the lateral force depends on the details of near-field coupling. If we recover the symmetry of the environment by adding another substrate above the sphere, the $y$-components of the Poynting vectors (Fig. 4c) and the SAM (Fig. 4f) become symmetrical and the lateral force vanishes again as numerically verified in Fig. 4g (red line).

The results here show that the lateral force emerges in a very general configuration of a chiral sphere above a substrate. The sign and magnitude of the lateral force are both directly related to the chirality of the particle. In the next section, we show with the help of an analytical theory that the lateral force is related to the chirality-generated lateral photon momentum.

**Lateral optical force on a dipolar chiral particle placed above a substrate.** The results shown up to now are full wave simulations but in order to obtain an intuitive understanding of the origin of the lateral force, we consider the configuration of a dipolar chiral particle above a substrate in which case the lateral optical force can be analytically evaluated and we will show that it comes from the lateral radiation pressure and spin density force arising from reflection. The induced dipole moments of such a chiral particle can be expressed as follows[28]:

$$\begin{bmatrix} \mathbf{p} \\ \mathbf{m} \end{bmatrix} = \begin{bmatrix} \alpha_{ee} & i\alpha_{em} \\ -i\alpha_{em} & \alpha_{mm} \end{bmatrix} \begin{bmatrix} \mathbf{E} \\ \mathbf{H} \end{bmatrix}, \qquad (2)$$

where **p** and **m** are the electric and magnetic moments, respectively; **E** and **H** are the fields acting on the particle. Here the polarizability of the particle is specified by parameters $\alpha_{ee}, \alpha_{mm}, \alpha_{em}$. We note that $\alpha_{em}$ is related to the chirality parameter $\kappa$ of the material that the particle is made of and $\alpha_{em}$ will changes sign if $\kappa$ changes sign and $\alpha_{em} = 0$ corresponds to a non-chiral particle.

The optical force acting on a dipolar chiral particle in an EM field can be derived (see Supplementary Note 1) based on the expression of the force acting on a dipolar particle [49-51],
$\langle \mathbf{F} \rangle = 1/2 \operatorname{Re} \left[ (\nabla \mathbf{E}^*) \cdot \mathbf{p} + (\nabla \mathbf{H}^*) \cdot \mathbf{m} - ck_0^4 / (6\pi)(\mathbf{p} \times \mathbf{m}^*) \right]$, where $\mathbf{E}$ and $\mathbf{H}$ are the fields acting on the particle. The force expression can be written as

$$\langle \mathbf{F} \rangle = \nabla U + \sigma \frac{\langle \mathbf{S} \rangle}{c} - \operatorname{Im}[\alpha_{em}] \nabla \times \langle \mathbf{S} \rangle + c\sigma_e \nabla \times \langle \mathbf{L}_e \rangle + c\sigma_m \nabla \times \langle \mathbf{L}_m \rangle$$
$$+ \omega \gamma_e \langle \mathbf{L}_e \rangle + \omega \gamma_m \langle \mathbf{L}_m \rangle + \frac{ck_0^4}{12\pi} \operatorname{Im}\left[ \alpha_{ee} \alpha_{mm}^* \right] \operatorname{Im}\left[ \mathbf{E} \times \mathbf{H}^* \right],$$
(3)

where $U = 1/4(\operatorname{Re}[\alpha_{ee}]|\mathbf{E}|^2 + \operatorname{Re}[\alpha_{mm}]|\mathbf{H}|^2 - 2\operatorname{Re}[\alpha_{em}]\operatorname{Im}[\mathbf{H} \cdot \mathbf{E}^*])$ is the term due to particle-field interaction; $\langle \mathbf{S} \rangle = 1/2 \operatorname{Re}[\mathbf{E} \times \mathbf{H}^*]$ is the time-averaged Poynting vector; $\langle \mathbf{L}_e \rangle$ and $\langle \mathbf{L}_m \rangle$ are the time-averaged spin densities defined before; $\sigma_e = k_0 / \varepsilon_0 \operatorname{Im}[\alpha_{ee}]$, $\sigma_m = k_0 / \mu_0 \operatorname{Im}[\alpha_{mm}]$ and $\sigma = \sigma_e + \sigma_m - c^2 k_0^4 / (6\pi)(\operatorname{Re}[\alpha_{ee}\alpha_{mm}^*] + \alpha_{em}\alpha_{em}^*)$ are the cross sections; $\gamma_e = -2\omega \operatorname{Im}[\alpha_{em}] + ck_0^4 / (3\pi\varepsilon_0) \operatorname{Re}[\alpha_{ee}\alpha_{em}^*]$ and $\gamma_m = -2\omega \operatorname{Im}[\alpha_{em}] + ck_0^4 / (3\pi\mu_0) \operatorname{Re}[\alpha_{mm}\alpha_{em}^*]$ also have the dimension of a cross section.

The first term in equation (3) corresponds to the gradient force. The second term represents the radiation pressure. The third term is a "vortex" force determined by the energy flow vortex around the particle and the optical activity ($\alpha_{em}$). The fourth and fifth terms are scattering forces associated with the curl of the spin densities and are called curl-spin forces [52-56]. The sixth and seventh terms which we refer to as spin density forces as they are directly related to the spin densities.

For an isolated chiral particle acted upon by a linearly polarized plane wave, the gradient force and the "vortex" force vanish because the fields' amplitudes and the time-averaged Poynting vector are constants. The last term contributes nothing as $\text{Im}[\mathbf{E} \times \mathbf{H}^*] = 0$ for a plane wave. The spin density terms vanish as we are considering linear polarization. As a consequence, the expression of the force reduces to $\langle \mathbf{F} \rangle = \sigma \langle \mathbf{S} \rangle / c$, which is just a forward scattering force.

In the presence of a substrate, the fields in equation (3) consist of both the incident plane wave and the reflected fields; that is, $\mathbf{E} = \mathbf{E}_{inc} + \mathbf{E}_{ref}, \mathbf{H} = \mathbf{H}_{inc} + \mathbf{H}_{ref}$, where the reflected fields can be expressed as:

$$\mathbf{E}_{ref} = \mu_0 \omega^2 \bar{\bar{G}}_{ref} \cdot \mathbf{p} + i\omega \left( \nabla \times \bar{\bar{G}}_{ref} \right) \cdot \mathbf{m}, \tag{4}$$

$$\mathbf{H}_{ref} = \varepsilon_0 \omega^2 \bar{\bar{G}}_{ref} \cdot \mathbf{m} - i\omega \left( \nabla \times \bar{\bar{G}}_{ref} \right) \cdot \mathbf{p}. \tag{5}$$

Here $\bar{\bar{G}}_{ref}$ is the Green's function for reflection describing the effect of the substrate and it takes into account the reflected fields including the propagating and the evanescent components [42]. We now refer to equations (3)-(5) to reveal the origin of the lateral force in the presence of a substrate. In equation (3), no lateral component can come from the gradient force as the free energy does not change when we apply a virtual in-plane (*xy*-plane) displacement of the particle. For a lossless particle under long wavelength condition, its polarizabilities are dominated by their real parts and as a consequence, the strengths of the vortex force and the last term are much smaller than that of the spin density force. For example, the strengths of these two terms are respectively about $10^{-6}$ and $10^{-3}$ smaller than that of the spin density force for a dipolar particle $(\varepsilon_r = 2, \mu_r = 1, \kappa = 1)$ of radius $a = 30$ nm placed at $d = 60$ nm above a semi-infinite gold substrate under the excitation of *z*-polarized plane wave $(\lambda = 600 \, \text{nm})$. We can hence focus on the radiation pressure term and the terms related to spin densities. In the presence of a substrate, the lateral component of the Poynting vector becomes asymmetric (Fig. 4b) and hence contributes to a lateral radiation pressure. The magnetic spin density is small compared with the contribution of its electric counterpart because the non-magnetic substrate gives a relatively stronger response to the electric field. Hence the total spin density can be written as $\langle \mathbf{L} \rangle = \langle \mathbf{L}_e \rangle + \langle \mathbf{L}_m \rangle \approx \langle \mathbf{L}_e \rangle \approx \varepsilon_0/(4\omega i) \mathrm{Im}[\mathbf{E}_{inc} \times \mathbf{E}_{ref}^*]$. As the incident electric field $\mathbf{E}_{inc}$ is along the *z* direction, this expression indicates that the direction of the spin density is mainly parallel to the surface of the substrate (as the numerical results shown in Fig. 4e confirm). Such a spin density can only contribute to a lateral force through the spin density terms but not through the curl-spin terms.

As the spin density is mainly attributed to $\langle \mathbf{L}_e \rangle$, the lateral component of the spin density force can be written as $F_y \approx \omega \gamma_e \hat{y} \cdot \langle \mathbf{L}_e \rangle \approx (\gamma_e \varepsilon_0/2)\hat{y} \cdot \mathrm{Im}[\mathbf{E}_{inc} \times \mathbf{E}_{ref}^*]$. Since the incident electric field is polarized along the *z* direction, the lateral force requires an *x*-component reflected field. Such a component exists only when there is a substrate and it emerges from the interference between the particle and the substrate that enables the cross-coupling of different Cartesian components of the scattered fields. This can be understood by directly examining the expression of the reflected field. Consider the excitation of the chiral particle under the incident magnetic field $H_{inc}$ which is along *y* direction. According to equation (2), an electric dipole moment $p^y = i\alpha_{em} H_{inc}$ can be induced, which produces the reflected field $E_{ref}^y = i\mu_0 \alpha_{em} \omega^2 G_{ref}^{yy} H_{inc}$ in the presence of a substrate. Here $G_{ref}^{yy}$ is the corresponding tensor element of the Green's function for reflection. The reflected field $E_{ref}^y$ further induces a *y*-direction magnetic dipole moment $m^y = -i\alpha_{em} E_{ref}^y$ and this magnetic dipole moment produces the required *x*-component electric field: $E_{\text{ref}}^x = i\mu_0\omega^3\alpha_{\text{em}}^2 G_{\text{ref}}^{yy}(\nabla\times\bar{\bar{G}}_{\text{ref}})^{xy}H_{\text{inc}}$. Note also that $\langle\mathbf{L}_e\rangle \propto \text{Im}[\mathbf{E}_{\text{inc}}\times\mathbf{E}_{\text{ref}}^*] \propto \alpha_{\text{em}}^2$ and hence the spin direction remains unchanged when $\alpha_{\text{em}}$ changes sign. Consequently, the sign of the lateral spin-density force is determined by the sign of the coefficient $\gamma_e$ as $\gamma_e \propto \alpha_{\text{em}}$.

Consider a spherical dipolar chiral particle ($a = 30$ nm) characterized by $\varepsilon_r = 2.0$, $\mu_r = 1.0$ and a chirality parameter $\kappa$ and the particle is placed at $d = 60$ nm above a semi-infinite gold substrate. One can derive the polarizability tensor elements $\alpha_{\text{ee}}, \alpha_{\text{mm}}$ and $\alpha_{\text{em}}$ for the particle from its Mie scattering coefficients (see Supplementary Note 2). The lateral force contributed by the radiation pressure and spin-density forces can then be analytically evaluated with the help of equation (3) (see Methods). Figure 5a shows the lateral force as a function of $\kappa$ at wavelength $\lambda = 600$ nm. It is clear that the lateral force vs. chirality relationship is similar to those calculated using full wave simulations for the bigger chiral sphere as shown in Fig. 4g. The analytical theory can also explain the oscillation phenomenon of the lateral force in the gold helix configuration (Fig. 2d). Let us now replace the gold substrate by a dielectric substrate ($\varepsilon_d = 2.5 + 0.001i$). We show in Fig. 5b the dependence of the lateral force on the thickness of the dielectric substrate according to the dipole theory. As expected, the force undergoes Fabry-Perot oscillations induced by the reflectance of the finite-thickness substrate. The dielectric substrate reflects the scattered field of the particle and causes constructive or destructive interference. A better understanding can be obtained by examining the reflection coefficient of the dielectric substrate described by the expression

$R = [2i(\zeta^2-1)\sin k_z d]/[(\zeta+1)^2 e^{-ik_z d} - (\zeta-1)^2 e^{ik_z d}]$, where $\zeta^{\text{TE}} = k_z/(\mu_d k_{0z})$, $\zeta^{\text{TM}} = k_z/(\varepsilon_d k_{0z})$ and $k_z^2 + k_{//}^2 = \varepsilon_d\mu_d k_0^2$. Here $k_{//}$ is the wave vector component parallel to the surface of the substrate. For a highly evanescent channel ($k_{//} \gg k_0$), $R^{\text{TE}} = 0$ and $R^{TM} \approx (\varepsilon_d^2 - 1)(1 - e^{2k_{//}t}) / [(\varepsilon_d - 1)^2 - (\varepsilon_d + 1)^2 e^{2k_{//}t}]$, where $R^{TM}$ is a monotonic function that increases with $t$ and takes a limit value of $(\varepsilon_d - 1)/(\varepsilon_d + 1)$. The red line in Fig. 5c shows the magnitude $|R^{TM}|$ as a function of $t$ for the case of $k_{//} = 10k_0$, where we see the reflection coefficient is a constant for large $t$. Propagating channels ($k_{//} < k_0$), on the other hand, behave quite differently. Figure 5c shows the $|R^{TE}|$ (green dotted line) and $|R^{TM}|$ (blue solid line) for the case of $k_{//} = 0.5k_0$. Clearly they undergo oscillations with a period that meets the condition $k_z \Delta t = \pi$ (here $\lambda = 600$ nm), indicating the oscillations are due to the Fabry-Perot resonances. Combining the properties of both the evanescent and propagating channels, the final reflected field should oscillate and the resulting lateral force should too.

## Discussion

While the lateral optical force is always there if the particle is chiral, it is clear that the lateral force is reasonably strong relative to the forward force only if the chirality is strong. The case with the helical gold particle, with numerical results show in Fig. 2, is a good example in which the maximum value of the lateral force can reach about 0.4 times that of the forward scattering force if the gold helix is put above a gold substrate. We note that if the incident wavelength is much larger than the helix dimension, the lateral force would become very weak because the total scattering cross section is small and in addition, the incident EM wave could not "resolve" the geometric handedness detail of the helix which is crucial for inducing the chiral effect[57]. Although we have only shown that a linearly polarized plane wave can induce a lateral force on a chiral particle, the force also exists when the plane wave is circularly polarized (see Supplementary Fig. 3 for the case of gold helices on gold substrate). The key point here is the asymmetric coupling of a chiral particle with a substrate, which can also be induced if the incident light is the circularly polarized. In addition, the lateral force also can be induced when the incident plane wave propagates in a direction perpendicular to the axis of the gold helix (see Supplementary Fig. 4).

In summary, we have numerically shown that the EM near-field coupling can induce a lateral optical force on a chiral particle near a substrate. The anomalous force can laterally push particles with opposite chirality in opposite directions. Using full-wave simulations, we established the existence of this counter-intuitive phenomenon for the helix-above-substrate configuration as well as the case of a chiral sphere with a simple constitutive relationship above a substrate. In the former case, the lateral force takes opposite signs for an LH and an RH helix. In the latter case, the lateral force changes sign when the chirality parameter changes sign from $+\kappa$ to $-\kappa$. The asymmetric coupling between a chiral particle and a substrate breaks the left-right symmetry due to the special "handedness" distribution of the Poynting vector (Fig.4b). By analytically deriving and evaluating the lateral force acting on a dipolar chiral particle above a substrate, we show that such lateral force must exist even in the small particle regime, and the force is attributed to the lateral component of the radiation pressure and "spin-density force" generated by the reflection field in presence of the substrate. The study reported here may find applications in the detection of chirality and in the separation of chiral molecules/enantiomers [60] using light induced forces. We note that the lateral force here must be accompanied by a torque induced on the chiral particles, the magnitude of which can be enhanced by a nearby surface. This may find applications in some opto-mechanical systems and as a light-driven "rotor" or "motor" [61]. However, we note that thermodynamic fluctuation effects and hydrodynamic properties of the environment should also be considered in such applications.

## Methods

**Numerical evaluation of optical forces by the Maxwell stress tensor method.** The optical forces acting on the gold helix and the chiral sphere are numerically evaluated using the Maxwell stress tensor [58], which is defined as

$$\bar{\bar{T}} = \varepsilon_0 \left[ \mathbf{EE} + c^2 \mathbf{BB} - \frac{1}{2}\left(\mathbf{E}\cdot\mathbf{E} + c^2 \mathbf{B}\cdot\mathbf{B}\right)\bar{\bar{I}} \right], \tag{6}$$

where $\bar{\bar{I}}$ is the unit tensor. To calculate the forces, we first obtain the EM total field using a commercial finite-element-method package COMSOL Multiphysics[45]. Then the tensor in equation (6) is integrated on a closed surface surrounding the particle. The forces shown in the main text are the time-averaged results.

**Semi-analytical evaluation of optical forces by dipole theory.** When a spherical dipolar chiral particle is placed above a substrate, the fields of the induced dipoles will be reflected by the substrate and react on the particle. Hence, the induced dipoles can be expressed as

$$\begin{bmatrix} \mathbf{p} \\ \mathbf{m} \end{bmatrix} = \begin{bmatrix} \alpha_{ee} & i\alpha_{em} \\ -i\alpha_{em} & \alpha_{mm} \end{bmatrix} \begin{bmatrix} \mathbf{E}_{inc} + \mathbf{E}_{ref} \\ \mathbf{H}_{inc} + \mathbf{H}_{ref} \end{bmatrix}. \tag{7}$$

For a small particle, the dipole polarizabilities in equation (7) can be related to the material parameters $\varepsilon_r, \mu_r, \kappa$ of the particle (see equation (1)) through Mie scattering theory (see Supplementary Note 2 and Ref. [40]).

Combining equation (7) with equations (4) and (5), the reflected fields can be written as

$$\begin{aligned} \mathbf{E}_{ref} &= \bar{\bar{\xi}}_1 \cdot \mathbf{E}_{inc} + \bar{\bar{\xi}}_2 \cdot \mathbf{H}_{inc}, \\ \mathbf{H}_{ref} &= \bar{\bar{\varsigma}}_1 \cdot \mathbf{E}_{inc} + \bar{\bar{\varsigma}}_2 \cdot \mathbf{H}_{inc}, \end{aligned} \tag{8}$$

where

$$\bar{\bar{\xi}}_1 = \left(\bar{\bar{D}}_1 \cdot \bar{\bar{C}}_1 + \bar{\bar{D}}_2\right) \cdot \left(\bar{\bar{I}} - \bar{\bar{D}}_1 \cdot \bar{\bar{C}}_1\right)^{-1}, \bar{\bar{\xi}}_2 = \left(\bar{\bar{D}}_1 \cdot \bar{\bar{C}}_2 + \bar{\bar{D}}_1\right) \cdot \left(\bar{\bar{I}} - \bar{\bar{D}}_1 \cdot \bar{\bar{C}}_1\right)^{-1},$$
$$\bar{\bar{\varsigma}}_1 = \left(\bar{\bar{C}}_1 \cdot \bar{\bar{D}}_2 + \bar{\bar{C}}_1\right) \cdot \left(\bar{\bar{I}} - \bar{\bar{C}}_1 \cdot \bar{\bar{D}}_1\right)^{-1}, \bar{\bar{\varsigma}}_2 = \left(\bar{\bar{C}}_1 \cdot \bar{\bar{D}}_1 + \bar{\bar{C}}_2\right) \cdot \left(\bar{\bar{I}} - \bar{\bar{C}}_1 \cdot \bar{\bar{D}}_1\right)^{-1},$$
(9)

and

$$\bar{\bar{C}}_1 = \left(-i\varepsilon_0\omega^2\alpha_{em}\bar{\bar{G}}_{ref} - i\omega\alpha_{ee}\nabla\times\bar{\bar{G}}_{ref}\right) \cdot \left(\bar{\bar{I}} - \varepsilon_0\omega^2\alpha_{mm}\bar{\bar{G}}_{ref} - \omega\alpha_{em}\nabla\times\bar{\bar{G}}_{ref}\right)^{-1},$$
$$\bar{\bar{C}}_2 = \left(\varepsilon_0\omega^2\alpha_{mm}\bar{\bar{G}}_{ref} + \omega\alpha_{em}\nabla\times\bar{\bar{G}}_{ref}\right) \cdot \left(\bar{\bar{I}} - \varepsilon_0\omega^2\alpha_{mm}\bar{\bar{G}}_{ref} - \omega\alpha_{em}\nabla\times\bar{\bar{G}}_{ref}\right)^{-1},$$
$$\bar{\bar{D}}_1 = \left(i\mu_0\omega^2\alpha_{em}\bar{\bar{G}}_{ref} + i\omega\alpha_{mm}\nabla\times\bar{\bar{G}}_{ref}\right) \cdot \left(\bar{\bar{I}} - \mu_0\omega^2\alpha_{ee}\bar{\bar{G}}_{ref} - \omega\alpha_{em}\nabla\times\bar{\bar{G}}_{ref}\right)^{-1},$$
$$\bar{\bar{D}}_2 = \left(\mu_0\omega^2\alpha_{ee}\bar{\bar{G}}_{ref} + \omega\alpha_{em}\nabla\times\bar{\bar{G}}_{ref}\right) \cdot \left(\bar{\bar{I}} - \mu_0\omega^2\alpha_{ee}\bar{\bar{G}}_{ref} - \omega\alpha_{em}\nabla\times\bar{\bar{G}}_{ref}\right)^{-1}.$$
(10)

The Green's function can be written as an integral in the momentum space as [42]

$$\bar{\bar{G}}_{ref} = \frac{i}{8\pi^2}\int_0^\infty dk_x dk_y \frac{1}{k_{0z}}\left[R^{TE}\hat{e}(k_{0z})e^{i\mathbf{k}_0\cdot\mathbf{r}}\hat{e}(-k_{0z})e^{-i\mathbf{k}_0'\cdot\mathbf{r}'} + R^{TM}\hat{h}(k_{0z})e^{i\mathbf{k}_0\cdot\mathbf{r}}\hat{h}(-k_{0z})e^{-i\mathbf{k}_0'\cdot\mathbf{r}'}\right].$$ (11)

Here $\mathbf{k}_0 = k_x\hat{x} + k_y\hat{y} + k_{0z}\hat{z}$, $\mathbf{k}_0' = k_x\hat{x} + k_y\hat{y} - k_{0z}\hat{z}$, $\hat{e}(k_{0z}) = \hat{e}(-k_{0z}) = (\hat{x}k_y - \hat{y}k_x)/k_{//}$,

$\hat{h}(k_{0z}) = -(\hat{x}k_x + \hat{y}k_y)k_{0z}/(k_0 k_{//}) + k_{//}\hat{z}/k_0$, $\hat{h}(-k_{0z}) = (\hat{x}k_x + \hat{y}k_y)k_{0z}/(k_0 k_{//}) + k_{//}\hat{z}/k_0$ and

$k_{//} = \sqrt{k_x^2 + k_y^2}$. The reflection coefficients $R^{TE}$ and $R^{TM}$ are defined in the main text.

In the calculations, $\bar{\bar{G}}_{ref}$ and its curl are first numerically evaluated and then one can apply equations (8)-(10) to evaluate the reflected fields $\mathbf{E}_{ref}$ and $\mathbf{H}_{ref}$. Once the fields acting on the particle are obtained, one can then apply the analytic results in equation (3) to calculate the radiation pressure and spin density forces which are the source of the lateral force.

**Numerical simulation.** All the full-wave EM simulations are performed with the package COMSOL Multiphysics. The relative permittivity of gold is described by a Drude model: $\varepsilon_{Au} = 1 - \omega_p^2/(\omega^2 + i\omega\omega_t)$, where $\omega_p = 1.37\times10^{16}$ rad s$^{-1}$ and $\omega_t = 4.084\times10^{13}$ rad s$^{-1}$ [59]. For the simulation of the gold helix systems we set $l\times w\times t = 4$ μm$\times 4$ μm$\times 200$ nm and for the case of the dielectric substrate we set $l \times w \times t = 4\ \mu m \times 2.4\ \mu m \times 200\ nm$. We choose a relatively larger substrate in the former case to reduce the effect caused by the reflection (at the edge) of the surface plasmons. For the simulation of the chiral sphere system, we set $l \times w \times t = 2\ \mu m \times 2\ \mu m \times 200\ nm$. The finite substrate does not affect the physics here; the phenomenon also exists in the case of an infinite substrate.

## Acknowledgements

This work was supported by AoE/P-02/12 and M-HKUST601/12. We thank Profs. Z. Q. Zhang and J. Ng and Dr. K. Ding for their valuable comments and suggestions.

## Author contributions

C.T.C. developed the concept and S.B.W. did the calculations. They wrote the paper together.

## Competing financial interests

The authors declare no competing financial interests.## Figure legends:

**Figure 1 | Anomalous lateral force in the helix particle-substrate configuration.** (**a**) A linearly polarized plane wave pushes a normal spherical particle forward. A standalone RH helix (**c**) and a standalone LH helix (**e**) will be pushed forward by radiation pressure. When the normal spherical particle is put near a substrate, (**b**) it still moves forward while the helical particles [(**d**) and (**f**)] are shifted in opposite directions by an anomalous lateral force. The arrows associated with the helical particles indicate the handedness. The substrate has the dimensions of $l \times w \times t$. (**g**) Dimensions of the helix particle. It has inner radius $r = 50$ nm, outer radius $R = 150$ nm and pitch $p = 300$ nm.

**Figure 2 | Numerically calculated lateral forces acting on the helical gold particles.** (**a**) Lateral forces $F_y$ acting on the LH (blue lines) and RH (red lines) particles in the presence of metal (gold) and dielectric ($\varepsilon_d = 2.5$) substrates. (**b**) Lateral forces as a function of the number of pitches for the LH (blue lines) and RH (red lines) particles in the presence (circles) and absence

of (triangles) the gold substrate. (**c**) Lateral force acting on the LH particle as a function of the gap distance between the particle and the gold substrate. (**d**) Lateral force as a function of the thickness of the dielectric substrate for the LH particle. In the numerical simulations, the frequency is set at $f = 490$ THz for case of the metallic substrate [(**b**) and (**c**)] and at $f = 380$ THz for the case of the dielectric substrate.

**Figure 3 | Magnetic field distribution on the gold substrate.** (**a**) $H_x$ field pattern for the RH particle on the gold substrate. (**b**) $H_x$ field pattern for the LH particle on the gold substrate. Note the asymmetrical pattern and the different directions in which the RH and LH particles scatter EM fields. The plotted fields are on a plane just above the substrate and the frequency is set at $f = 420$ THz.

**Figure 4 | Numerical results for a chiral sphere above a gold substrate**. Time-averaged Poynting vectors (**a**) for an isolated chiral sphere, (**b**) for a chiral sphere above a gold substrate and (**c**) for a chiral sphere sandwiched symmetrically by two gold substrates. We set $\kappa = 1$ for the sphere. Time-averaged electric spin density $\langle \mathbf{L}_e \rangle$ (**d**) for the isolated sphere, (**e**) for the sphere above a gold substrate and (**f**) for the sphere sandwiched by two gold substrates. The left-right asymmetry is obvious in panels (b) and (e). (**g**) Lateral force acting on the chiral sphere above a gold substrate (blue line) and sandwiched by two gold substrates (red line) as a function of the chirality parameter $\kappa$. The blue line shows that the sign of the lateral force $F_y$ depends on $\kappa$ and $F_y = 0$ if $\kappa = 0$. The red line indicates the lateral force vanishes in the sandwiched case. The frequency is set at $f = 500$ THz.

**Figure 5 | Analytical results for a dipolar chiral particle above a substrate.**
(**a**) Lateral force acting on a dipolar chiral particle ($a = 30$ nm, $\varepsilon_r = 2.0$) as a function of its chirality when the particle is located 60 nm above a semi-infinite gold substrate. (**b**) Lateral force acting on the chiral particle ($\kappa = -1$) as a function of the thickness of a dielectric substrate ($\varepsilon_d = 2.5 + 0.001i$), showing oscillating behavior. (**c**) Magnitudes of the reflection coefficients for an evanescent channel $k_{//} = 10k_0$ (red line) and a propagating channel $k_{//} = 0.5k_0$ (green dotted line and blue line). The wavelength is set to be $\lambda = 600$ nm.

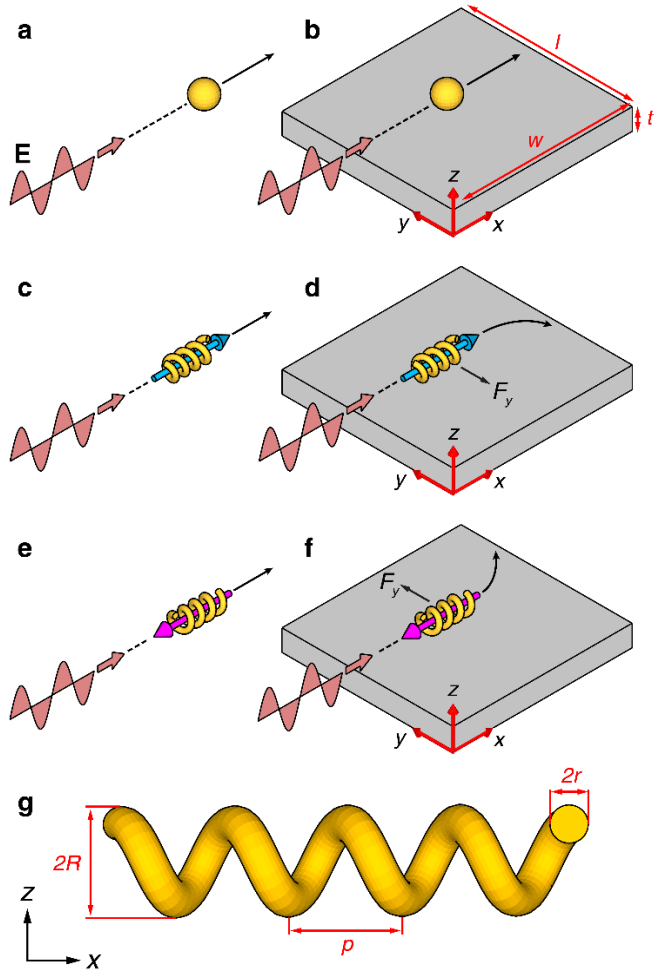

**Figure 1**

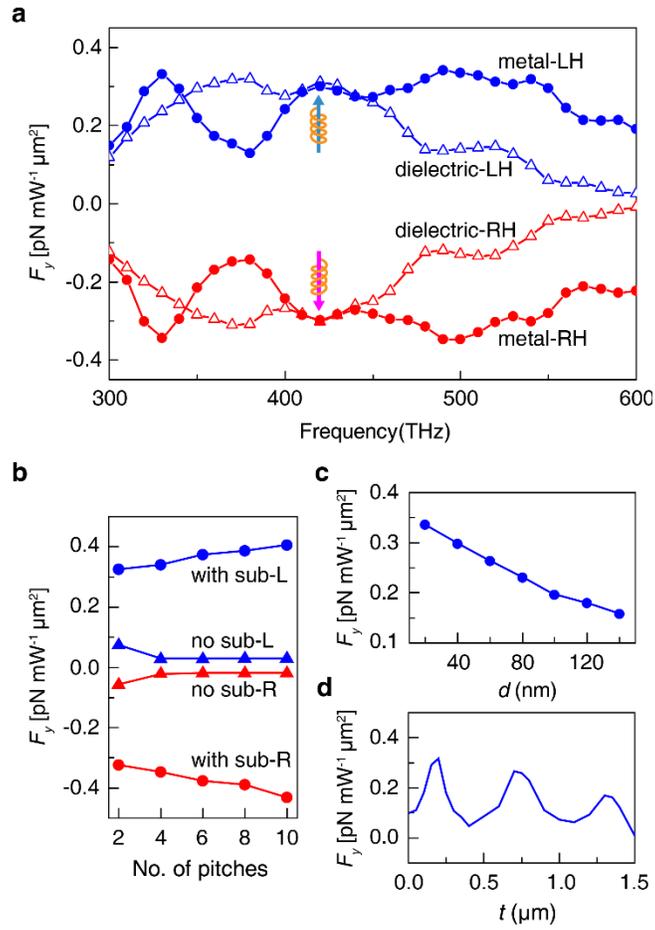

**Figure 2**

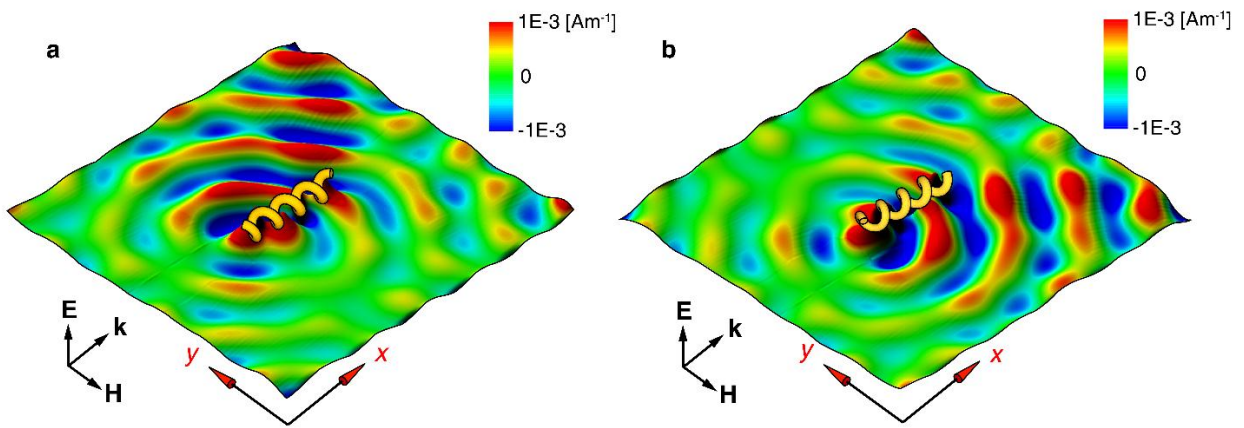

**Figure 3**

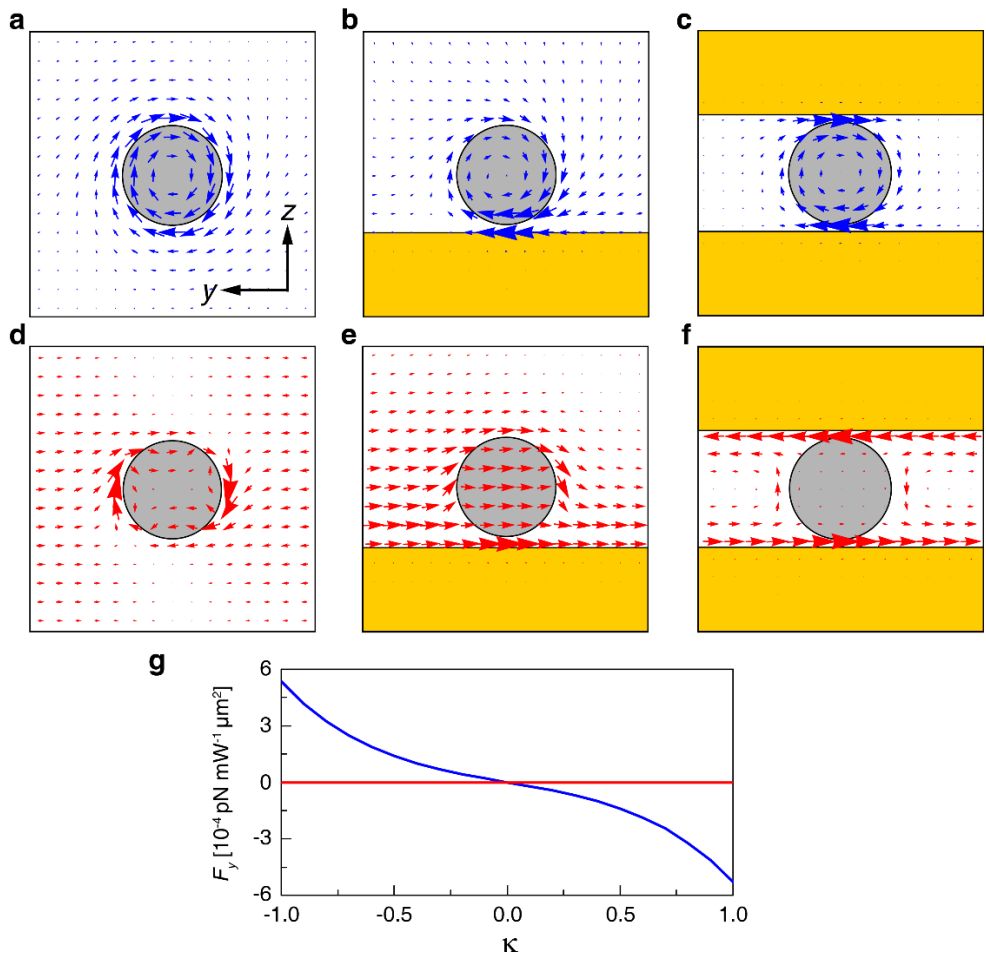

**Figure 4**

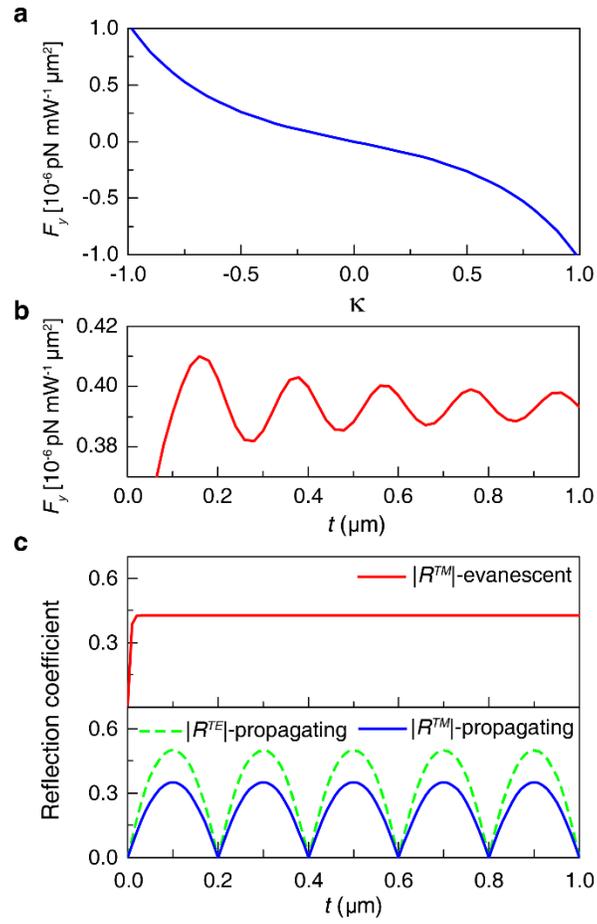

**Figure 5**

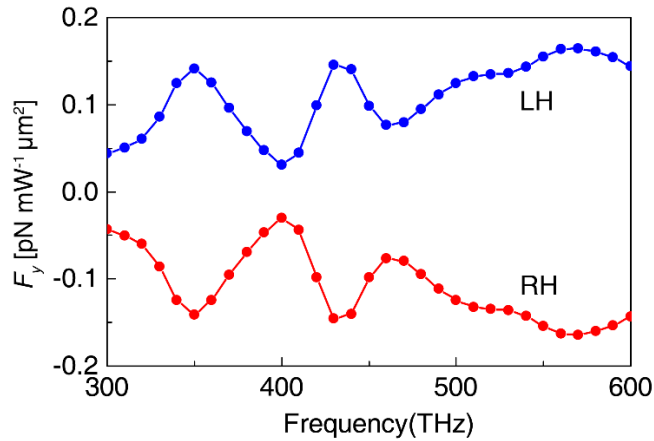

**Supplementary Figure 1 | Lateral forces acting on the gold helices above a perfect-electric-conductor substrate**. The configurations of the substrate and the helices are the same as specified in the main text. The blue line corresponds to the lateral optical force of the left-handed helix while the red line corresponds to the lateral optical force of the right-handed helix.

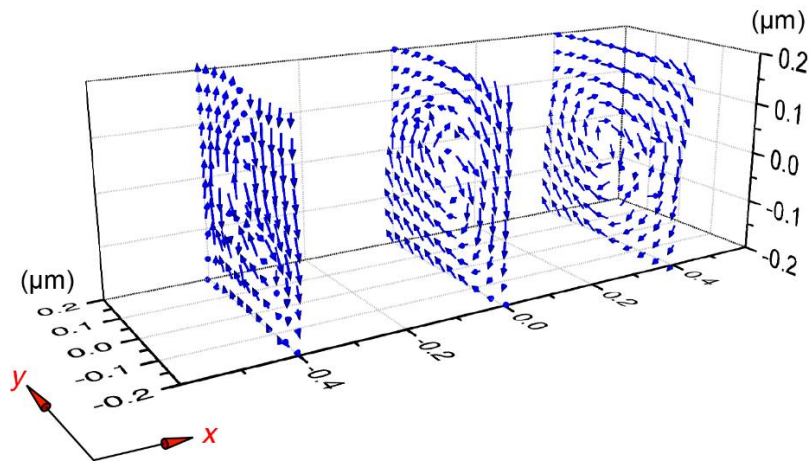

**Supplementary Figure 2 | Time-averaged Poynting vectors projected on three xz planes in the case of an isolated right-handed gold helix.** The gold helix has the same dimension as specified in the main text. The wave vector of the incident plane wave is along *x* direction and the electric field polarization is along *z* direction. The helix axis is along the *x* direction passing through the origin of the coordinates. The Poynting vectors circulate around the helix, very similar to what we show in Fig. 4a for a chiral sphere.

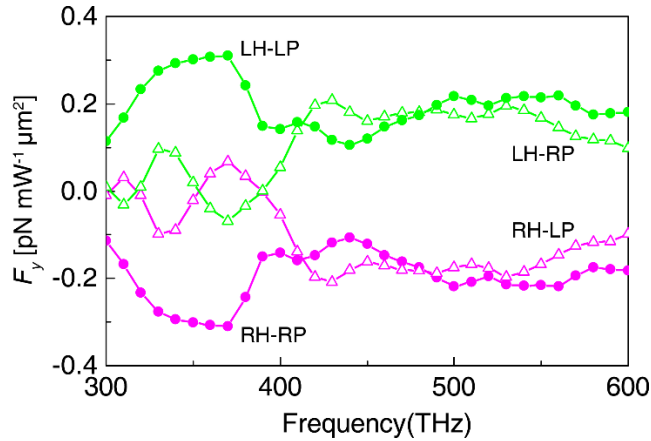

**Supplementary Figure 3 | Lateral forces acting on the gold helices above a gold substrate under the excitation of circularly polarized plane waves.** The configurations of the substrate and the helices are the same as specified in the main text. The wave vector of the incident wave is along the axis of the helices. We considered four cases here: left-handed helix under the incidence of a plane wave with left-handed circular polarization (LH-LP), left-handed helix under the incidence of a plane wave with right-handed circular polarization (LH-RP), right-handed helix under the incidence of a plane wave with left-handed circular polarization (RH-LP), right-handed helix under the incidence of a plane wave with right-handed circular polarization (RH-RP).

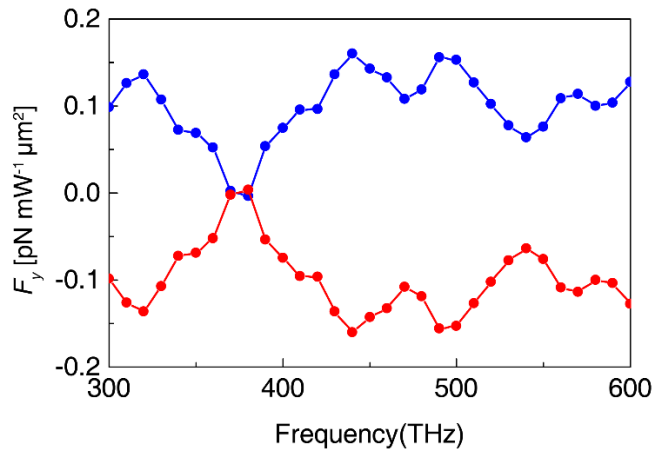

**Supplementary Figure 4 | Lateral forces acting on the gold helices above a gold substrate under the excitation of a linearly polarized plane wave with wave vector perpendicular to the axis of the helices.** The configurations of the substrate and the helices are the same as specified in the main text. The wave vector of the incident plane wave is along $y$ direction and the electric field polarization is along $z$ direction. The blue line corresponds to the lateral force of the left-handed helix while the red line corresponds to the lateral force of the right-handed helix. The results show that the optical lateral force can also be induced if the incident wave vector is perpendicular to the axis of the helix.

## Supplementary Note 1

**Optical force acting on a dipolar chiral particle induced by a general beam field**

Here, we limit ourselves to particles that can be described by equation (2) in the main text. A more general derivation can be found in Ref. [1]. The vector expression of the optical force acting on a dipole particle writes

$$\langle \mathbf{F} \rangle = \frac{1}{2} \mathrm{Re} \left[ (\nabla \mathbf{E}^*) \cdot \mathbf{p} + (\nabla \mathbf{H}^*) \cdot \mathbf{m} - \frac{ck_0^4}{6\pi} (\mathbf{p} \times \mathbf{m}^*) \right] \tag{1}$$

where $\mathbf{E}$ and $\mathbf{H}$ are the electric and magnetic fields that act on the particle. Substitute equation (2) of the main text into the above equation (1), we obtain

$$\langle \mathbf{F} \rangle = \frac{1}{2} \mathrm{Re} \begin{bmatrix} \alpha_{ee}(\nabla \mathbf{E}^*) \cdot \mathbf{E} + \alpha_{mm}(\nabla \mathbf{H}^*) \cdot \mathbf{H} + i\alpha_{em}(\nabla \mathbf{E}^*) \cdot \mathbf{H} - i\alpha_{em}(\nabla \mathbf{H}^*) \cdot \mathbf{E} \\ -\frac{ck_0^4}{6\pi}(\alpha_{ee}\mathbf{E} + i\alpha_{em}\mathbf{H}) \times (\alpha_{mm}^*\mathbf{H}^* + i\alpha_{em}^*\mathbf{E}^*) \end{bmatrix}. \tag{2}$$

Note that we have neglected the subscript "inc" here. Let us address the terms on the right-hand side of equation (2) one by one. The first term can be expressed as

$$\begin{aligned}
&\mathrm{Re}\left[\alpha_{ee}(\nabla \mathbf{E}^*) \cdot \mathbf{E}\right] \\
&= \mathrm{Re}\left\{\alpha_{ee}\left[(\mathbf{E} \cdot \nabla)\mathbf{E}^* + \mathbf{E} \times (\nabla \times \mathbf{E}^*)\right]\right\} \\
&= \mathrm{Re}[\alpha_{ee}]\left\{\frac{1}{2}\left[(\mathbf{E} \cdot \nabla)\mathbf{E}^* + (\mathbf{E}^* \cdot \nabla)\mathbf{E}\right] + \mathrm{Re}\left[\mathbf{E} \times (\nabla \times \mathbf{E}^*)\right]\right\} \\
&\quad -\frac{1}{2i}\mathrm{Im}[\alpha_{ee}]\left[(\mathbf{E} \cdot \nabla)\mathbf{E}^* - (\mathbf{E}^* \cdot \nabla)\mathbf{E}\right] + \omega\mu_0 \mathrm{Im}[\alpha_{ee}]\mathrm{Re}[\mathbf{E} \times \mathbf{H}^*] \\
&= \frac{1}{2}\mathrm{Re}[\alpha_{ee}]\nabla|\mathbf{E}|^2 + 2\omega\mu_0 \mathrm{Im}[\alpha_{ee}]\langle \mathbf{S}\rangle + \frac{2\omega}{\varepsilon_0}\mathrm{Im}[\alpha_{ee}]\nabla \times \langle \mathbf{L}_e\rangle,
\end{aligned} \tag{3}$$

Similarly,

$$\mathrm{Re}\left[\alpha_{mm}(\nabla \mathbf{H}^*) \cdot \mathbf{H}\right] = \frac{1}{2}\mathrm{Re}[\alpha_{mm}]\nabla|\mathbf{H}|^2 + 2\omega\varepsilon_0 \mathrm{Im}[\alpha_{mm}]\langle \mathbf{S}\rangle + \frac{2\omega}{\mu_0}\mathrm{Im}[\alpha_{mm}]\nabla \times \langle \mathbf{L}_m\rangle, \tag{4}$$

For the cross coupling terms we have

$$\text{Re}\left[i\alpha_{em}(\nabla \mathbf{E}^*)\cdot\mathbf{H} - i\alpha_{em}(\nabla \mathbf{H}^*)\cdot\mathbf{E}\right]$$

$$= \text{Re}\left\{i\alpha_{em}\left[(\mathbf{H}\cdot\nabla)\mathbf{E}^* + \mathbf{H}\times(\nabla\times\mathbf{E}^*) - (\mathbf{E}\cdot\nabla)\mathbf{H}^* - \mathbf{E}\times(\nabla\times\mathbf{H}^*)\right]\right\}$$

$$= -\text{Re}[\alpha_{em}]\text{Im}\left[(\mathbf{H}\cdot\nabla)\mathbf{E}^* - (\mathbf{E}\cdot\nabla)\mathbf{H}^*\right] - \text{Im}[\alpha_{em}]\text{Im}[\omega\mu_0\mathbf{H}\times\mathbf{H}^* + \omega\varepsilon_0\mathbf{E}\times\mathbf{E}^*] \quad (5)$$

$$- \text{Im}[\alpha_{em}]\text{Re}\left[(\mathbf{H}\cdot\nabla)\mathbf{E}^* - (\mathbf{E}\cdot\nabla)\mathbf{H}^*\right],$$

$$= -\text{Re}[\alpha_{em}]\text{Im}\left[\nabla(\mathbf{H}\cdot\mathbf{E}^*)\right] - 4\omega^2\text{Im}[\alpha_{em}]\left(\langle\mathbf{L}_m\rangle + \langle\mathbf{L}_e\rangle\right) - 2\text{Im}[\alpha_{em}]\nabla\times\langle\mathbf{S}\rangle$$

The last part of equation (2) can be expressed as

$$-\frac{ck_0^4}{6\pi}\text{Re}\left[(\alpha_{ee}\mathbf{E} + i\alpha_{em}\mathbf{H})\times(\alpha_{mm}^*\mathbf{H}^* + i\alpha_{em}^*\mathbf{E}^*)\right]$$

$$= -\frac{ck_0^4}{6\pi}\left\{\begin{array}{l}\text{Re}[\alpha_{ee}\alpha_{mm}^*]\text{Re}[\mathbf{E}\times\mathbf{H}^*] - \text{Im}[\alpha_{ee}\alpha_{mm}^*]\text{Im}[\mathbf{E}\times\mathbf{H}^*] \\ +\text{Re}[\alpha_{ee}\alpha_{em}^*](i\mathbf{E}\times\mathbf{E}^*) - \text{Re}[\alpha_{em}\alpha_{mm}^*](i\mathbf{H}\times\mathbf{H}^*) - \alpha_{em}\alpha_{em}^*\text{Re}[\mathbf{H}\times\mathbf{E}^*]\end{array}\right\} \quad (6)$$

$$= -\frac{ck_0^4}{6\pi}\left\{\begin{array}{l}2\left(\text{Re}[\alpha_{ee}\alpha_{mm}^*] + \alpha_{em}\alpha_{em}^*\right)\langle\mathbf{S}\rangle - \dfrac{4\omega}{\varepsilon_0}\text{Re}[\alpha_{ee}\alpha_{em}^*]\langle\mathbf{L}_e\rangle \\ -\dfrac{4\omega}{\mu_0}\text{Re}[\alpha_{em}\alpha_{mm}^*]\langle\mathbf{L}_m\rangle - \text{Im}[\alpha_{ee}\alpha_{mm}^*]\text{Im}[\mathbf{E}\times\mathbf{H}^*]\end{array}\right\}$$

Combining (3)-(6), we write the final expression of the force as

$$\langle\mathbf{F}\rangle = \nabla U + \sigma\frac{\langle\mathbf{S}\rangle}{c} - \text{Im}[\alpha_{em}]\nabla\times\langle\mathbf{S}\rangle + c\sigma_e\nabla\times\langle\mathbf{L}_e\rangle + c\sigma_m\nabla\times\langle\mathbf{L}_m\rangle$$
$$+ \omega\gamma_e\langle\mathbf{L}_e\rangle + \omega\gamma_m\langle\mathbf{L}_m\rangle + \frac{ck_0^4}{12\pi}\text{Im}[\alpha_{ee}\alpha_{mm}^*]\text{Im}[\mathbf{E}\times\mathbf{H}^*], \quad (7)$$

where

$$U = \frac{1}{4}\text{Re}[\alpha_{ee}]|\mathbf{E}|^2 + \frac{1}{4}\text{Re}[\alpha_{mm}]|\mathbf{H}|^2 - \frac{1}{2}\text{Re}[\alpha_{em}]\text{Im}[\mathbf{H}\cdot\mathbf{E}^*]$$

$$\sigma_e = \frac{k_0\text{Im}[\alpha_{ee}]}{\varepsilon_0}, \sigma_m = \frac{k_0\text{Im}[\alpha_{mm}]}{\mu_0}, \sigma = \sigma_e + \sigma_m - \frac{c^2k_0^4}{6\pi}\left(\text{Re}[\alpha_{ee}\alpha_{mm}^*] + \alpha_{em}\alpha_{em}^*\right). \quad (8)$$

$$\gamma_e = -2\omega\text{Im}[\alpha_{em}] + \frac{ck_0^4}{3\pi\varepsilon_0}\text{Re}[\alpha_{ee}\alpha_{em}^*], \gamma_m = -2\omega\text{Im}[\alpha_{em}] + \frac{ck_0^4}{3\pi\mu_0}\text{Re}[\alpha_{mm}\alpha_{em}^*]$$

## Supplementary Note 2

**Polarizability of a spherical dipolar chiral particle**

Assume that the incident plane wave that impinges a spherical chiral particle has an amplitude $E_0$ and is propagating in $z$ direction and its electric field is polarized along $x$ direction. Following the notations of Bohren and Huffman, the electric field generated by the particle can be expressed as

$$E_{sca} = \sum_{n=1} E_n \left[ i a_n N_{e1n} - b_n M_{o1n} + c_n M_{e1n} - i d_n N_{o1n} \right],  \quad (9)$$

where $E_n = E_0 i^n (2n+1)/[n(n+1)]$ and $N_{e1n}, M_{o1n}, M_{e1n}, N_{o1n}$ are the vector harmonics and $a_n$, $b_n$, $c_n$, $d_n$ are the scattering coefficients defined as:

$$a_n = \frac{V_n(R) A_n(L) + V_n(L) A_n(R)}{W_n(L) V_n(R) + W_n(R) V_n(L)},$$

$$b_n = \frac{W_n(L) B_n(R) + W_n(R) B_n(L)}{W_n(L) V_n(R) + W_n(R) V_n(L)}, \quad (10)$$

$$c_n = -d_n = i \frac{W_n(R) A_n(L) - W_n(L) A_n(R)}{W_n(L) V_n(R) + W_n(R) V_n(L)}$$

and

$$W_n(J) = m \psi_n(m_J x) \xi'_n(x) - \xi_n(x) \psi'_n(m_J x),$$
$$V_n(J) = \psi_n(m_J x) \xi'_n(x) - m \xi_n(x) \psi'_n(m_J x),$$
$$A_n(J) = m \psi_n(m_J x) \psi'_n(x) - \psi_n(x) \psi'_n(m_J x), \quad (11)$$
$$B_n(J) = \psi_n(m_J x) \psi'_n(x) - m \psi_n(x) \psi'_n(m_J x).$$

Here $J = L, R$ and $x = k_0 a$. $\psi_n(\rho) = \rho j_n(\rho)$, $\xi_n(\rho) = \rho h_n^{(1)}(\rho)$ with $j_n(\rho)$ being the spherical Bessel functions and $h_n^{(1)}(\rho)$ being the spherical Hankel functions of the first kind. Assume the surrounding medium is vacuum, the relative refractive indices $m_L, m_R$ and the mean refractive index $m$ take the expressions: $m_L = \sqrt{\varepsilon_r \mu_r} + \kappa$, $m_R = \sqrt{\varepsilon_r \mu_r} - \kappa$, $m = (m_L + m_R)/2$.

Within the small particle limit, we retain only the $n = 1$ term and equation (9) can be written as

$$E_{sca} \approx E_1 \left[ i a_1 N_{e11} - b_1 M_{o11} + c_1 M_{e11} - i d_1 N_{o11} \right]. \quad (12)$$

In the far-field limit, the vector harmonics can be approximated as:

$$N_{e11} \approx -i(\cos\theta\cos\phi\hat{\theta} - \sin\phi\hat{\phi})\frac{e^{ik_0r}}{k_0r}, \quad N_{o11} \approx -i(\cos\theta\sin\phi\hat{\theta} + \cos\phi\hat{\phi})\frac{e^{ik_0r}}{k_0r},$$
$$M_{e11} \approx (\sin\phi\hat{\theta} + \cos\theta\cos\phi\hat{\phi})\frac{e^{ik_0r}}{k_0r}, \quad M_{o11} \approx (-\cos\phi\hat{\theta} + \cos\theta\sin\phi\hat{\phi})\frac{e^{ik_0r}}{k_0r}. \quad (13)$$

Substitute (13) into (12), we have the scattered field written as

$$E_{sca} \approx i\frac{3}{2}E_0 \frac{e^{ik_0r}}{k_0r}\left[\begin{array}{l}a_1(\cos\theta\cos\phi\hat{\theta} - \sin\phi\hat{\phi}) + b_1(\cos\phi\hat{\theta} - \cos\theta\sin\phi\hat{\phi}) \\ +c_1(\sin\phi\hat{\theta} + \cos\theta\cos\phi\hat{\phi}) - d_1(\cos\theta\sin\phi\hat{\theta} + \cos\phi\hat{\phi})\end{array}\right]. \quad (14)$$

At the same time, the electric fields in the far-field generated by the electric dipole and magnetic dipole induced on a dipolar chiral particle are given by

$$\lim_{r\to\infty} E_{sca}^{p} = \frac{k_0^3}{4\pi\varepsilon_0}\frac{e^{ik_0r}}{k_0r}(\hat{r}\times\vec{p})\times\hat{r} = \frac{k_0^3}{4\pi\varepsilon_0}\frac{e^{ik_0r}}{k_0r}\hat{r}\times(\alpha_{ee}E_0\hat{x} + i\alpha_{em}Z_0^{-1}E_0\hat{y})\times\hat{r} \quad (15)$$

$$\lim_{r\to\infty} E_{sca}^{m} = -\frac{Z_0 k_0^3}{4\pi\mu_0}\frac{e^{ik_0r}}{k_0r}\hat{r}\times(\mu_0\vec{m}) = -\frac{Z_0 k_0^3}{4\pi\mu_0}\frac{e^{ik_0r}}{k_0r}\hat{r}\times(-i\alpha_{em}E_0\hat{x} + \alpha_{mm}Z_0^{-1}E_0\hat{y}) \quad (16)$$

where $Z_0 = \sqrt{\mu_0/\varepsilon_0}$ is the impedance of vacuum and we have used the expressions: $\mathbf{p} = \alpha_{ee}\mathbf{E} + i\alpha_{em}\mathbf{H}$, $\mathbf{m} = -i\alpha_{em}\mathbf{E} + \alpha_{mm}\mathbf{H}$. Substitute $\hat{x} = \sin\theta\cos\phi\hat{r} + \cos\theta\cos\phi\hat{\theta} - \sin\phi\hat{\phi}$ and $\hat{y} = \sin\theta\sin\phi\hat{r} + \cos\theta\sin\phi\hat{\theta} + \cos\phi\hat{\phi}$ into the above equations and the total scattered fields can be written as:

$$\lim_{r\to\infty} E_{sca}^{pm} = E_0 \frac{e^{ik_0r}}{k_0r}\left[\begin{array}{l}\frac{k_0^3}{4\pi\varepsilon_0}\alpha_{ee}(\cos\theta\cos\phi\hat{\theta} - \sin\phi\hat{\phi}) + \frac{ik_0^3 c}{4\pi}\alpha_{em}(\cos\theta\sin\phi\hat{\theta} + \cos\phi\hat{\phi}) \\ +\frac{ik_0^3 c}{4\pi}\alpha_{em}(\sin\phi\hat{\theta} + \cos\theta\cos\phi\hat{\phi}) + \frac{k_0^3}{4\pi\mu_0}\alpha_{mm}(\cos\phi\hat{\theta} - \cos\theta\sin\phi\hat{\phi})\end{array}\right] \quad (17)$$

Comparing equation (17) with (14), we obtain the following expressions for the polarizability tensor elements of the dipolar chiral particle:

$$\alpha_{ee} = \frac{i6\pi\varepsilon_0}{k_0^3}a_1, \quad \alpha_{mm} = \frac{i6\pi\mu_0}{k_0^3}b_1, \quad \alpha_{em} = \frac{6\pi}{k_0^3 c}c_1 = -\frac{6\pi}{k_0^3 c}d_1 \quad (18)$$

## Supplementary Reference